# Using a hydrogen-bond index to predict the gene-silencing efficiency of siRNA based on the local structure of mRNA

Kathy Q. Luo[1,2] and Donald C. Chang[1*]

[1] Department of Biology and [2] Department of Chemical Engineering, the Hong Kong University of Science and Technology, Clear Water Bay, Hong Kong, China

*Key words:* RNA interference, siRNA, gene silencing, mRNA structure, H-bond.

The gene silencing effect of short interfering RNA (siRNA) is known to vary strongly with the targeted position of the mRNA. A number of hypotheses have been suggested to explain this phenomenon. We would like to test if this positional effect is mainly due to the secondary structure of the mRNA at the target site. We proposed that this structural factor can be characterized by a single parameter called "the hydrogen bond (H-b) index", which represents the average number of hydrogen bonds formed between nucleotides in the target region and the rest of the mRNA. This index can be determined using a computational approach. We tested the correlation between the H-b index and the gene-silencing effects on three genes (Bcl-2, hTF and cyclin B1) using a variety of siRNAs. We found that the gene-silencing effect is inversely dependent on the H-b index, indicating that the local mRNA structure at the targeted site is the main cause of the positional effect. Based on this finding, we suggest that the H-b index can be a useful guideline for future siRNA design.

## INTRODUCTION

RNA interference (RNAi) is an evolutionarily conserved mechanism for repressing targeted gene expression [1]. It exerts its silencing effect by mediating sequence-specific mRNA degradation. In this process, double-stranded RNA (dsRNA) molecules are cleaved by a ribonuclease (Dicer) into 21-23 bp fragments called "short interfering RNA" (siRNA) [2-4]. The siRNA in turn binds to a protein complex called "RNA-induced silencing complex (RISC)", and targets it to mRNAs that have complementary sequence with the siRNA [5-7]. Then, the targeted mRNAs are destroyed through cleavage by RISC [8].

Application of siRNA has recently become an important tool for suppressing the expression of specific genes. The efficiency of gene silencing, however, varied significantly between siRNAs targeted to different positions of a gene [4,9-12]. At present, there is still a lack of clear understanding on the mechanisms that determine the gene silencing efficiency of a given siRNA. A number of hypotheses have been proposed in the literature, including: **(a)** Local protein factor(s) on the mRNA may cause the positional effect [10]. **(b)** The local structure of the targeted mRNA may affect the accessibility of the siRNA [7,11,12]. **(c)** Factors such as sequence-dependent mRNA product release or differential efficiency of 5' siRNA phosphorylation may influence the efficacy of the siRNA [13]. Among these different proposals, we think the structural factor may be the most important one. Two earlier comparative studies between siRNA and antisense oligonucleotides (ASO) had suggested that the local structure of the mRNA had a strong effect on the suppression of gene expression [7, 12]. One of these studies utilized scanning oligonucleotide arrays to screen ASO that hybridized strongly to the mRNA [12]. The local structure was not directly visualized in this case. In the other study, siRNAs targeted to two local structures (a stem and a loop) were analysed using a computational approach [7]. Other more general structures have not yet been examined.

---

*Corresponding author. Email: bochang@ust.hk

We have two major objectives in this study. First, we would like to find a quantitative parameter that can characterize the accessibility of the siRNA to the targeted mRNA based on a structural consideration. Second, we would like to test the structural hypothesis by examining the correlation between the gene-silencing effect and the accessibility of the siRNA in a large variety of local mRNA structures. The efficiency of mRNA degradation by the RISC should be dependent on accessibility of the siRNA to the target region of the mRNA. Our hypothesis is that, since nucleotides in the mRNA can often form hydrogen bonds (i.e., becoming double-stranded) with other nucleotides in the same mRNA molecule, if the target region of the mRNA has a more loosen structure (i.e. less hydrogen bonding), it will be easier for the siRNA to bind with the targeted mRNA through base-pairing. This means that the mRNA accessibility can be quantified using an H-b index (see below). Thus, if the positional effect of siRNA is determined mainly by the local structure of mRNA at the target site, siRNAs with a lower H-b index should have a higher efficiency in suppressing gene expression. Otherwise, alternative causes such as local protein factors may be more important. Here, we reported results of several experimental tests using three different genes, including Bcl-2, hTF and cyclin B.

## MATERIALS AND METHODS

*1. Preparation of siRNA duplexes*

21-mer RNA oligonucleotides were synthesized by Dharmacon Inc. (Lafayette, CO). siRNA duplexes in the 2'-deprotected and desalted form were dissolved in a 1X universal buffer (provided by the compony) in concentration of 20 μM.

*2. Determination of mRNA structure*

The possible structures of a mRNA molecule was obtained using the "Mfold" web server, which provides several closely related softwares for predicting the secondary structure of single stranded nucleic acids [14]. The gateway for the Mfold web server is http://www.bioinfo.rpi.edu/ applications/mfold. The accession numbers of genes used for Mfold analysis are the following: human Bcl-2 (AX057146), human Tissue Factor (hTF) (M16553) and human cyclin B1 (NM_031966).

*3. Electroporation and Western blotting*

The siRNA was introduced into cells using an electroporation method based on our previously study [15]. HeLa cells were grown in Minimum Essential Medium (MEM) supplemented with 10% fetal bovine serum (FBS). Cells were then trypsinised and washed once with poration medium (PM), which contained 260 mM mannitol; 5 mM sodium phosphate; 10 mM potassium phosphate; 1 mM $MgCl_2$ and 10 mM HEPES (pH 7.3). A proper amount of cells resuspended in PM was added with 2-4 μl of rhodamine-dextran (10 kDa) (1 mM), 8-10 μl of siRNA duplexes (20 μM) and 3 μg of cyclin B-GFP plasmid DNA (only in the co-transfection experiments) to make up a total volume of 100 μl. The cell mixture was transferred into a Gene Pluser® Cuvette (0.1 cm electrode) (Bio-Rad) and electroporated using the GENE PULSER® II RF Module (Bio-Rad) under the conditions: 120 V and 10 RF pulses with duration of 2.2 ms. Following electroporation, cells were transferred into a 100-mm culture plate containing MEM plus 10% FBS and cultured at 37°C in a $CO_2$ incubator. After 24 to 36 hrs, cells were collected and analyzed using the standard Western blot technique. The antibodies used included: anti-Bcl2 (c-2) monoclonal antibody (Santa Cruz); anti-ß tubulin (654162) monoclonal antibody (Calbiochem); anti-cyclin B1 (GNS-11) monoclonal antibody (BD PharMingen); anti-Cdc2 monoclonal antibody (Santa Cruz) and anti-GFP polyclonal antibody (Molecular Probes). Reactive bands were detected using the ECL™ system (Amersham) and their intensity was measured using the MetaMorph software (Universal Imaging, West Chester, PA).

*4. Imaging techniques*

After cyclin B-GFP plasmid (with or without siRNA) was introduced into HeLa cells by electroporation, cells were grown on an observation chamber that contained a glass coverslip at the bottom. After 24 or 36 hours, cells were washed twice with culture medium. Their fluorescent images were then recorded using a fluorescence microscope (Axiovert 35, Zeiss) equipped with a cooled CCD camera (MicroMax, by Princeton Instruments) and processed digitally using the MetaMorph software.

## RESULTS

## 1. The gene silencing efficiency of a siRNA is dependent on its targeting position in the gene

To demonstrate that the gene silencing effect of RNAi depends on the targeting region, we synthesized two siRNA duplexes against different regions of Bcl-2 gene. The siRNA duplex was introduced into HeLa cells using an electroporation method [15]. The effect of siRNA on Bcl-2 protein level was evaluated 48 hrs later by Western blot analysis using antibody against Bcl-2 (Fig. 1A). We found that the siRNA of Bcl-2-N (which targets to the N-terminus of Bcl-2 between amino acids 51-69) reduced the Bcl-2 level to 39% of the control, while siRNA of Bcl-2-C (which targets to the C-terminus of Bcl-2 between amino acids 429-447) had little effect in reducing the level of Bcl-2 protein (Fig. 1B). No reduction of β-tubulin protein level was observed, suggesting that the silencing effect generated by the siRNA of Bcl-2-N was gene-specific. This result showed that siRNAs against different regions of Bcl-2 gene can have different silencing effects. We think that this result is due to the differences in the secondary structures of Bcl-2 mRNA at the targeted sites. Indeed, using the Mfold program [14], we found a loop structure in the mRNA at the region targeted by Bcl-2-N, while a hairpin structure was seen in the mRNA complementing to Bcl-2-C (Fig. 1C). The difference in their local structures may explain why the two siRNAs had different gene silencing efficiency.

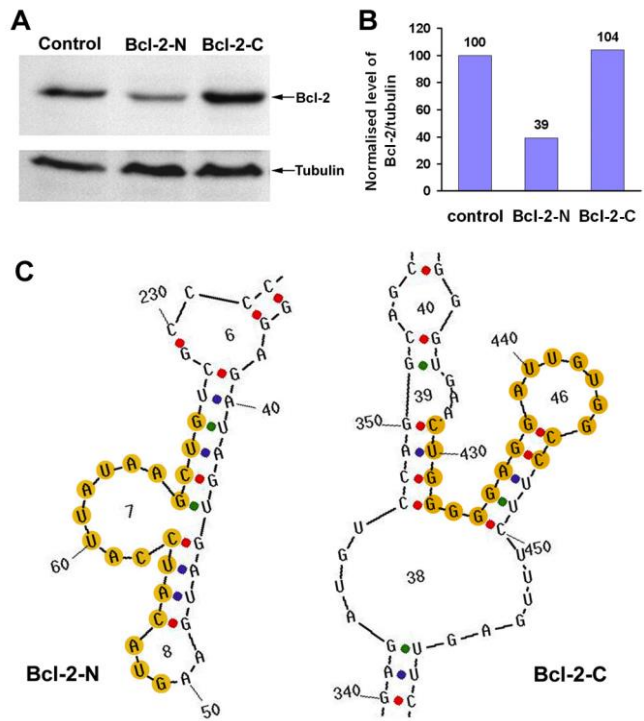

**Figure 1. Suppression of Bcl-2 gene expression.** (A) Western blot analysis on HeLa cells electroporated with or without siRNA duplexes Bcl-2-N or Bcl-2-C. (B) Based on results of the Western blot, the expression levels of the Bcl-2 gene (normalised by ß-tubulin protein level) were determined under conditions with or without the application of the siRNAs. (C) The local secondary structures of Bcl-2 mRNA at the regions targeted by the siRNAs Bcl-2-N and Bcl-2-C. The nucleotides targeted by siRNA are colored in orange. The RNA structures were generated using the Mfold software.

## 2. The secondary structure of the mRNA at the siRNA target region can be characterized by an H-b index

The above explanation, however, is not totally satisfactory. Since as many as 21 possible secondary structures can be generated for the Bcl-2 mRNA using the Mfold software, one cannot rely on a single predicted structure of the mRNA to explain the effect of a given siRNA. Instead, one needs to have a quantitative parameter that can characterize the over-all structural effects. Thus, we propose to introduce a parameter called the "H-b index", which represents the average number of hydrogen bonds formed in all possible secondary structures of Bcl-2 mRNA at the region targeted by the siRNA (as determined using a computational approach). The H-b index is defined by the following formula:

$$\textbf{H-b index} = \sum_{i=1}^{19} (\textbf{\textit{P}}_{H\text{-bond formation}} \times \textbf{\textit{N}}_{HB})_i \qquad (1)$$

Where, $i$ is the index representing each mRNA nucleotide in the region targeted by the siRNA, $\textbf{\textit{P}}_{H\text{-bond formation}}$ is the probability that the $i^{th}$ nucleotide can form H-bonds with other nucleotides within the same mRNA. $\textbf{\textit{P}}_{H\text{-bond formation}}$ is calculated based on all possible structures of a mRNA molecule predicted by the Mfold software, i.e.,

$$\textbf{\textit{P}}_{H\text{-bond formation}} = 1 - \frac{\text{Number of structures that the } i^{th} \text{ nucleotide is in a single strand}}{\text{Total number of possible structures of the mRNA molecule}} \qquad (2)$$

$N_{HB}$ is the number of hydrogen bonds that the $i^{th}$ nucleotide can form. $N_{HB}$ equals 3 for nucleotides of G or C, and 2 for A or U. The possible structures of a mRNA molecule was obtained using the "Mfold" web server, which provides several closely related softwares for predicting the secondary structure of single stranded nucleic acids [14]. The gateway for the Mfold web server is http://www.bioinfo.rpi.edu/ applications/mfold. Table 1 shows an example of calculating the H-b index for the Bcl-2-N siRNA based on the Mfold analysis of the human Bcl-2 mRNA.

**Table 1. Calculation of the H-b index for the siRNA Bcl-2-N based on 21 secondary structures of Bcl-2 mRNA predicted by the Mfold program**

| Position and nucleotide of siRNA | # of single strand structures[a] | Probability of H-bond formation[b] | Average # of H-bond formed |
|---|---|---|---|
| 51 G | 15 | 0.29 | 0.86 |
| 52 U | 15 | 0.29 | 0.57 |
| 53 A | 20 | 0.05 | 0.10 |
| 54 C | 2 | 0.90 | 2.71 |
| 55 A | 2 | 0.90 | 1.81 |
| 56 U | 1 | 0.95 | 1.90 |
| 57 C | 5 | 0.76 | 2.29 |
| 58 C | 12 | 0.43 | 1.29 |
| 59 A | 12 | 0.43 | 0.86 |
| 60 U | 12 | 0.43 | 0.86 |
| 61 U | 16 | 0.24 | 0.48 |
| 62 A | 16 | 0.24 | 0.48 |
| 63 U | 12 | 0.43 | 0.86 |
| 64 A | 15 | 0.29 | 0.57 |
| 65 A | 15 | 0.29 | 0.57 |
| 66 G | 2 | 0.90 | 2.71 |
| 67 C | 2 | 0.90 | 2.71 |
| 68 U | 10 | 0.52 | 1.05 |
| 69 G | 5 | 0.76 | 2.29 |
| % of GC= 36.8% | | | H-b index = 24.95 |

Note: (a) Number of RNA structures having the nucleotide being single stranded. (b) The definition of "probability of H-bond formation" is given in equation (2).

Using the same method, we have also calculated the H-b index for the siRNA of Bcl-2-C. The results, together with other relevant structural information, are summarized in Table 2. The H-b index of Bcl-2-N (24.95) was found to be significantly lower than that of Bcl-2-C (32.38). According to our hypothesis, a lower H-b index would indicate a better accessibility of the siRNA to the targeted mRNA. Thus, the observation that Bcl-2-N was more effective in silencing the Bcl-2 gene than Bcl-2-C is consistent with the prediction of our hypothesis.

Table 2. Properties of siRNAs targeted to the human Bcl-2 gene

| siRNA name | siRNA sequence | Position | RNA structure | % of GC | H-b index |
|---|---|---|---|---|---|
| Bcl-2-N | GUACAUCCAUUAUAAGCUG | 51-69 | loop | 36.8 | 24.95 |
| Bcl-2-C | CUGGGGGAGGAUUGUGGCC | 429-447 | hairpin | 68.4 | 32.38 |

Note: The H-b index was determined using eq. (1) and based on the secondary structures of Bcl-2 mRNA predicted by the Mfold program (14).

### 3. H-b index is highly correlated with the gene silencing efficiency of siRNA in human tissue factor gene

In order to test the correlation between the H-b index and the siRNA efficiency more extensively, we have analysed the positional effects of 14 different siRNAs in silencing the human tissue factor (hTF) gene based on a recent study reported by Holen et al [10]. Figure 2 shows the local secondary structures of the hTF mRNA in regions targeted by various siRNAs. (Here, siRNA-targeted nucleotides are highlighted in orange color). Based on structures generated from the Mfold software, we calculated the H-b index for every siRNA designed for the hTF gene (Table 3). Also, based on the experimental results reported by Holen et al [10], we calculated the gene silencing efficiency of each siRNA on both the endogenous hTF gene and the over-expressed hTF-luciferase (hTF-Luc) gene. These results are summarized in Table 3.

Table 3. Comparison of H-b index and efficiency of gene silencing

| siRNA name | siRNA sequence | RNA structure | % of GC | H-b index | % Reduction in hTF-Luc[a] | % Reduction in hTF[b] | Average reduction |
|---|---|---|---|---|---|---|---|
| 77i | uggagaccccugccuggcc | stem | 73.68 | 41.4 | 13 | 5 | 9.1 |
| 167i | gcgcuucaggcacuacaaa | 3 loops | 52.63 | 24.7 | 80 | 85 | 82.3 |
| 256i | cccgucaaucaagucuaca | stem | 47.37 | 35.1 | 71 | 43 | 57.0 |
| 372i | gaagcagacguacuuggca | 1 large loop | 52.63 | 18.0 | 84 | 76 | 80.1 |
| 459i | cuccccagaguucacaccu | stem | 57.89 | 41.4 | 22 | 2 | 12.0 |
| 478i | uaccuggagacaaaccucg | hairpin | 52.63 | 20.8 | 16 | 13 | 14.5 |
| 562i | cggacuuuagucagaagga | hairpin | 47.37 | 29.4 | 50 | 13 | 31.3 |
| 929i | gcuggaaggagaacucccc | hairpin | 63.16 | 28.3 | 19 | 5 | 11.9 |
| 158i | agguggccggcgcuucagg | stem | 73.68 | 36.5 | 27 | 0 | 13.5 |
| 161i | uggccggcgcuucaggcac | stem | 73.68 | 35.1 | 26 | 26 | 25.8 |
| 164i | ccggcgcuucaggcacuac | stem + loop | 68.42 | 29.0 | 58 | 71 | 64.3 |
| 167i | gcgcuucaggcacuacaaa | loops | 52.63 | 24.7 | 79 | 87 | 83.2 |
| 170i | cuucaggcacuacaaauac | loops | 42.11 | 20.2 | 59 | 34 | 46.6 |
| 173i | caggcacuacaaauacugu | loops | 42.11 | 22.4 | 68 | 68 | 67.9 |
| 176i | gcacuacaaauacuguggc | 1 large loop | 47.37 | 28.1 | 44 | 40 | 41.8 |

Note:(a) Data were based on a previous study reported by (10). (b) Reduction of endogenous hTF mRNA levels were measured based on northern blot analysis of hTF mRNA [Fig. 3A & 3B in (10)]. The mRNA levels were normalised to a loading control (GAPDH) and standardized to mock-transfected cells.

The data in Table 3 were organized into two groups: The first group consists of siRNAs targeted to different parts of the hTF gene, while the second group consists of overlapping siRNAs targeted to neighboring regions of one specific site in the hTF gene (Figs. 2B & 2I). From results shown in group #1, one can easily see that: (1) The gene silencing effects was not correlated with the order of position of the siRNA-targeting region. (2) All siRNAs containing a hairpin structure were generally not very effective in gene silencing. (3) For those siRNAs that had a significant effect of gene silencing, their H-b index values were relatively low.

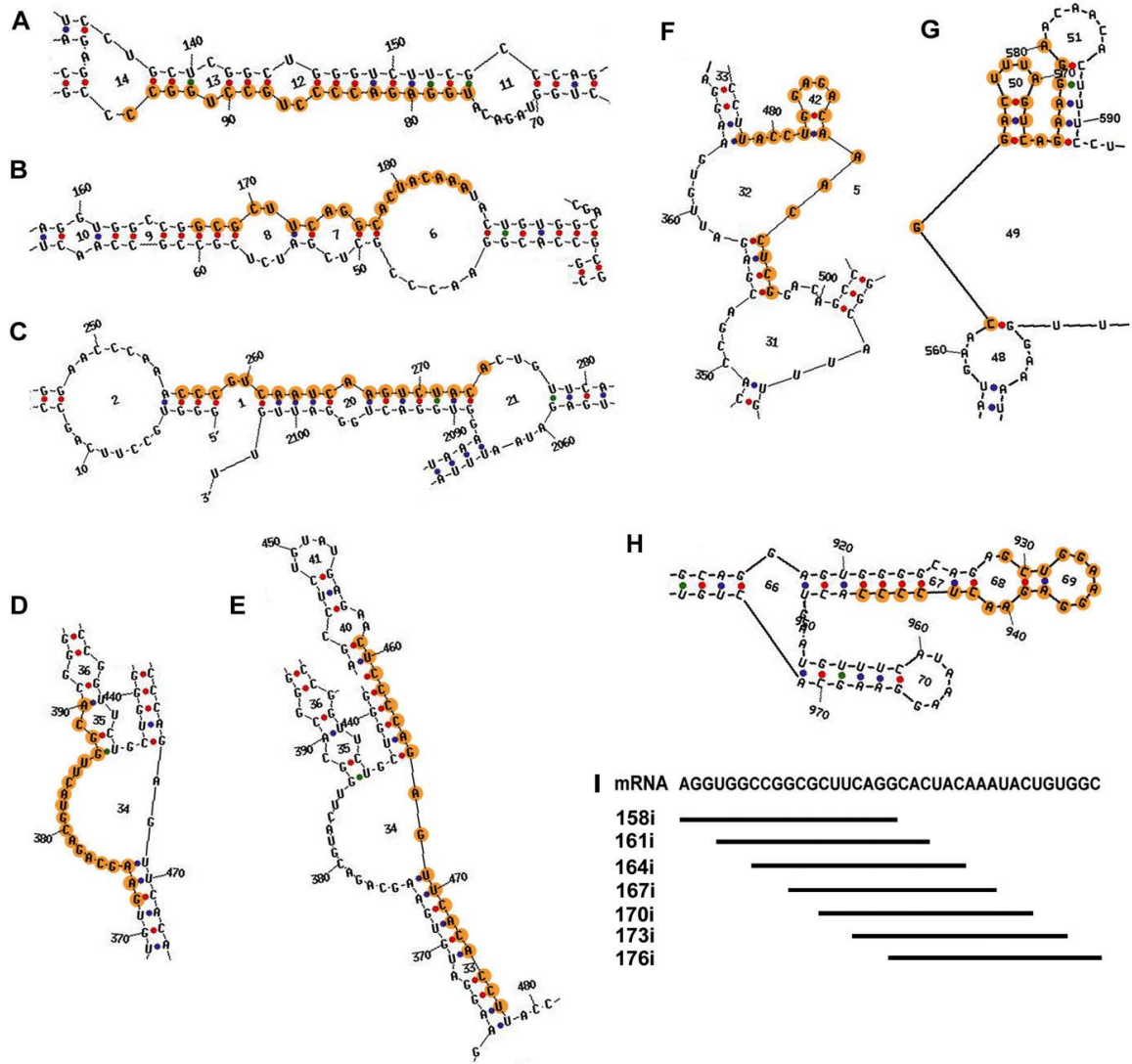

**Figure 2. Local secondary structures of hTF (human tissue factor) mRNA.** All local structures were from the predicted structure #1 using the Mfold program. Panels A-H shows the structures targeted by siRNA of 77i (A), 167i (B), 256i (C), 372i (D), 459i (E), 478i (F), 562i (G) and 929i (H). Here, the siRNA targeting sequence was colored in orange. Panel I shows the nucleotide sequence of hTF mRNA at the targeting site surrounding siRNA 158i-176i. Seven siRNAs were generated by shifting the nucleotide sequence in increments of 3. Their structures can be seen in Panel B.

Next, we conducted a more quantitative analysis by plotting the percentage of hTF gene reduction versus the H-b index for all non-hairpin siRNAs in group #1 (Fig. 3A). It gave an almost linear relationship (the coefficient of determination $R^2 = 0.879$). (Note: A perfect linear correlation would give $R^2 = 1.0$). One may question the validity of our analysis by pointing out that, since a siRNA having a higher % of GC content may in general tend to have a larger H-b index, the results shown in Fig. 3A may simply indicate that the gene-silencing effect is related to the GC content of the siRNA. Thus, for purpose of comparison, we have also plotted the percentage of hTF gene reduction versus the % of GC content of the same siRNAs (Fig. 3B). The correlation was far less clear ($R^2 = 0.407$). These results suggest that the efficiency of gene silencing for hTF had a higher correlation with the local mRNA structure than with the GC content.

This conclusion is further supported by data from group #2 of Table 3, where seven overlapping siRNAs were designed to target to one specific region of the hTF gene (Figs. 2B & 2I). We have examined the correlation between the percentage of reduction in the reporter gene (hTF-Luc) and the H-b index (Fig. 3C) or the % of GC content (Fig. 3D). Again, a higher linear regression ($R^2=0.706$) was found in the H-b index, while a lesser correlation ($R^2=0.449$) was seen between the effect of RNA interference and the GC content.

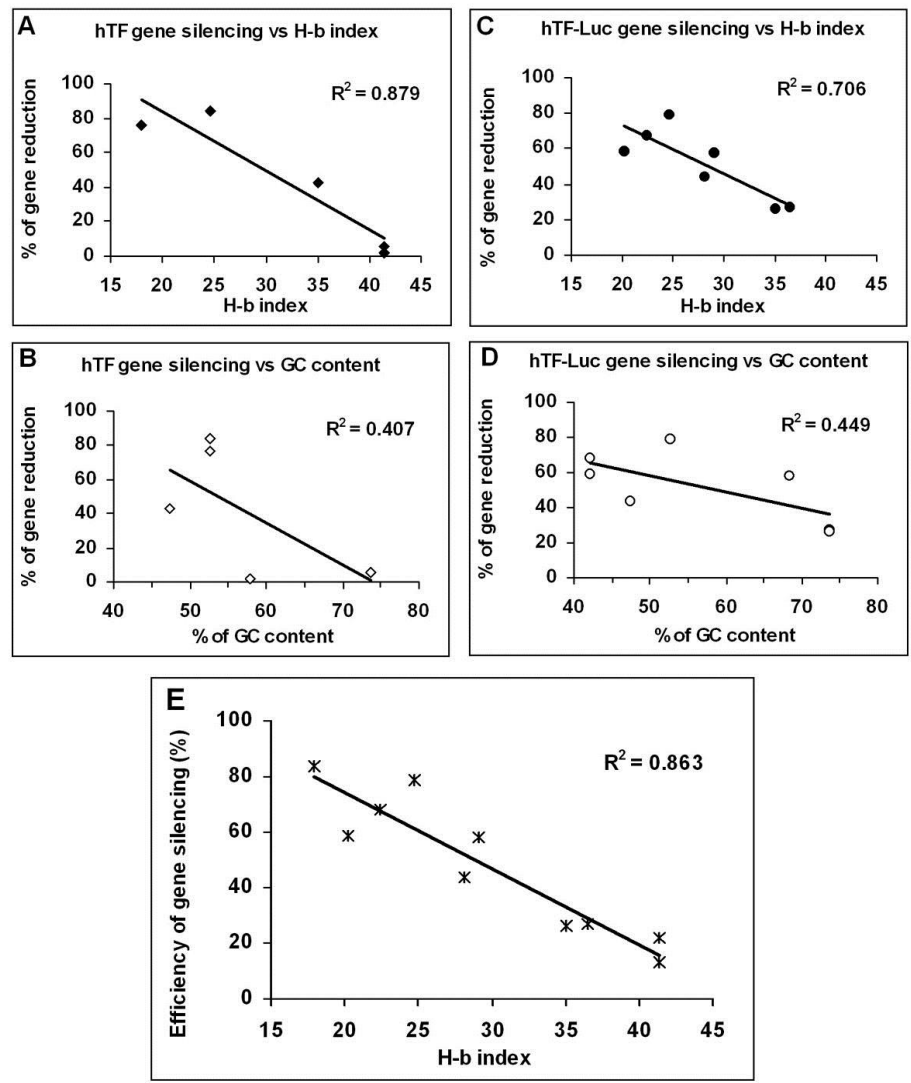

**Figure 3. Relationship between siRNA efficacy and the H-b index or GC content.** Using data summarized in Table 3, we analyzed the gene silencing effect vs the H-b index (A, C & E) or the GC content (B, D). The gene silencing efficiency was based on the measured reduction of expression either in the endogenous hTF gene (A, B) or in the hTF-Luc reporter gene (C, D & E). Panels A and B were based on data of non-hairpin siRNAs listed in group #1 of Table 3. Panels C and D were based on data of seven overlapping siRNA listed in group #2 of Table 3. Panel E shows the combined result from 10 non-hairpin siRNAs listed in Table 3.

Finally, we combined both Group #1 and Group #2 data and plotted the efficiency of gene silencing (based on % of suppression in expressing the hTF-Luc reporter gene) against the H-b index for ten siRNA samples that did not contain a hairpin structure* (77i, 158i, 161i, 164i, 167i, 170i, 173i, 176i, 372i and 459i) (Fig. 3E). We found a highly significant correlation between the gene silencing effect and the H-b index ($R^2 = 0.863$), indicating that the gene-silencing efficiency is largely dependent on the secondary structure of the mRNA at the target site. From these results, we concluded that: (1) siRNA having a lower H-b index in general would have a higher gene-silencing efficiency, and (2) siRNA forming a hairpin structure usually would be less effective in gene-silencing. (*Note: Fig. 3E did not include the data of siRNA 256i, which hybridized with both the 5' and 3' ends of the mRNA. Since regulatory proteins binding to these terminal regions may interfere with the local secondary structure, they could affect the siRNA-mRNA interaction at this target site).

### 4. Testing the prediction of our model using siRNAs against different regions of cyclin B1 gene

In order to further test the above observations, we have designed several siRNAs against the human cyclin B1 gene at different regions. Their properties are summarised in Table 4. These siRNAs were specially targeted to loop (L1, L2), stem (S1) or hairpin (H1) structures of the human cyclin B1 mRNA. These siRNAs have similar GC content but different H-b index values (L1<L2<S1). If our structural hypothesis is correct, L1 should have the highest efficiency in silencing the cyclin B1 gene, L2 should have an intermediate effect, and S1 and H1 should be least effective in gene silencing.

Table 4. Properties of siRNAs targeted to the human cyclin B1 gene

| siRNA name | siRNA sequence | Position | RNA structure | % of GC | H-b index |
| --- | --- | --- | --- | --- | --- |
| L1 | CAGCUACUGGAAAAGUCAU | 383-401 | loop | 42.1% | 17.06 |
| L2 | GAGCCAUCCUAAUUGACUG | 782-800 | loop | 47.4% | 25.06 |
| S1 | CCUGAGCCUAUUUUGGUUG | 526-544 | stem | 47.4% | 35.66 |
| H1 | AGCCCAAUGGAAACAUCUG | 559-577 | hairpin | 47.4% | 25.38 |

In this study, siRNA of L1, L2, S, or H1 was introduced into HeLa cells by electroporation. After 32 or 48 hrs, cell extracts were collected and analyzed by SDS-PAGE followed with Western blot analysis using antibody against cyclin B1 (Fig. 4A). The amount of cyclin B1 protein was determined quantitatively by measuring the intensity of protein bands of cyclin B which was normalized by the level of Cdc2 detected in the same protein blot. The normalised results are shown in Figure 4B. At 32 hrs after electroporation, application of the L1 siRNA was found to produce a very strong gene silencing effect; it reduced the expression level of the endogenous cyclin B1 gene to 6% of the control. By comparison, application of the L2 siRNA only reduced the cyclin B1 level to 25%. The siRNAs of S1 and H1, on the other hand, did not show any significant gene silencing activity. Similar results were obtained from cells collected at 48 hrs after applying the various siRNAs (data not shown). These results demonstrated that a low H-b index value was indeed correlated with a high gene-silencing effect, and siRNA containing a hairpin structure was generally ineffective.

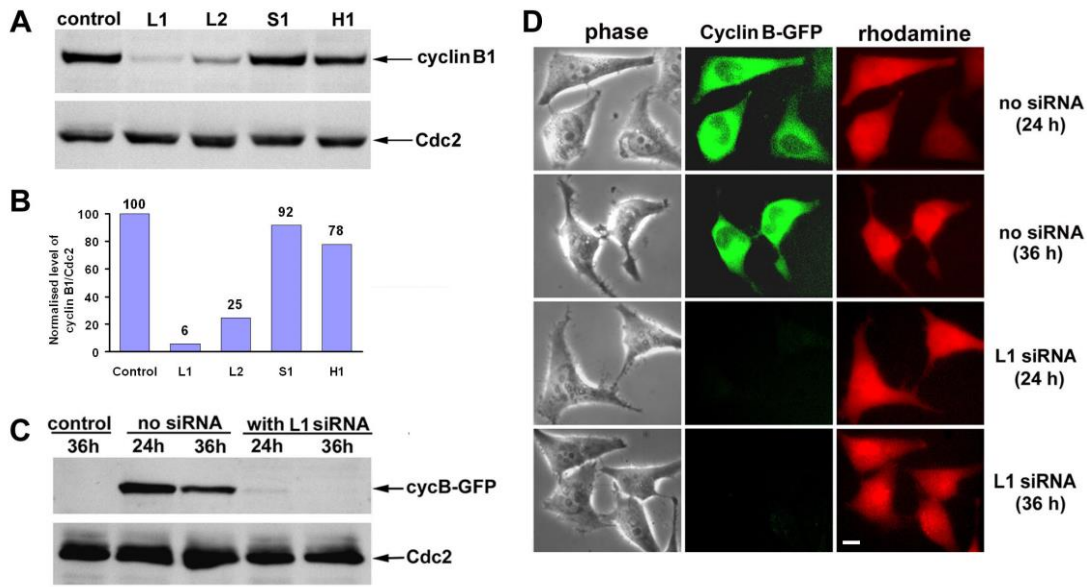

**Figure 4. Gene silencing effects of siRNAs on the endogenous and exogenous expression of cyclin B1.** (A) Western blot analysis of cyclin B1 protein in HeLa cells at 32 hrs after electroporation with various siRNA duplexes. Control cells were electroporated with buffer only. Protein levels of Cdc2 in the same samples were shown here as loading controls. (B) The relative levels of the expressed cyclin B1 under various conditions were determined and normalised by the levels of Cdc2. (C) The expressed level of cyclin B1-GFP (with or without L1 siRNA) at 24 and 36 hrs after gene transfection were analysed by Western blot analysis. Control sample had neither siRNA nor cyclin B1-GFP. (D) Fluorescence images showing the expression of cyclin B1-GFP with or without applying the L1 siRNA. The images for rhodamine indicated that all cells under study were successfully electroporated. Scale bar: 10 μm.

To further verify our experimental results, we have examined the effects of different siRNAs on repressing the expression of an exogenous cyclin B1-GFP fusion gene. Plasmid DNA containing the cyclin B1-GFP gene was introduced into HeLa cells by electroporation with or without the L1 siRNA. The efficiency of siRNA up-take was monitored by co-electroporation with rhodamine-dextran (10 kDa). Results of gene expression assays based on Western blot analysis are shown in Figure 4C. In the absence of L1 siRNA, cyclin B1–GFP was found to express at a high level at 24 and 36 hrs after electroporation. In the presence of L1 siRNA, the level of cyclin B1-GFP protein observed at 24 hrs reduced dramatically, and no cyclin B1-GFP was detected at 36 hrs.

Similar findings were also obtained using an imaging technique. As shown in Figure 4D, in the absence of L1 siRNA, a high level of cyclin B1-GFP fluorescent protein was observed in the transfected HeLa cells at either 24 or 36 hrs after electroporation. However, in cells co-electroporated with both cyclin B1-GFP and L1 siRNA, very little GFP fluorescence signal was detected. These results again demonstrated that the expression of cyclin B1-GFP gene was almost completely inhibited by the L1 siRNA (Fig. 4D).

## DISCUSSION

RNA interference is a powerful approach for studying gene function in many organisms. Thus, there is a strong interest to develop and improve this technique. At this time, a key question is how to design siRNAs that can have high efficiency in gene silencing. We know that the effectiveness of a siRNA is highly dependent on its target position [10,16]. The mechanism, however, was not clear. Some hypothesized that local protein factors on different regions of mRNA might cause the positional effect [10]. Others suggested that the activity of siRNA is mainly affected by the secondary structure of mRNA at the target site [7,11,12]. Although these two proposed mechanisms are not mutually exclusive, in this study, we found evidence suggesting that the local structure of mRNA at the target site is a dominant factor. Furthermore, we showed that one can use a single parameter, the H-b index, to characterize the overall structural effects. This information, we believe, can be greatly helpful in optimizing the design of effective siRNAs for silencing a specific gene.

The major advantage of using the H-b index is that one can bypass the cumbersome process of guessing the correct secondary structure of a mRNA molecule at local regions. At present, computer softwares can only predict a set of possible secondary structures for a given mRNA. For example, using the Mfold program, we can obtain 32 predicted RNA structures for the human cyclin B1 mRNA. It is difficult to know which one of the 32 possible structures represents the real folding of the mRNA in a cell. In this work, we proposed to use a statistical approach to solve this problem. By introducing the concept of the H-b index, we can use a single parameter to reflect the overall probability for nucleotides within the siRNA targeting region to form double stranded complex with other parts of the mRNA. A low value of the H-b index would mean that most of the nucleotides within the target region are in single stranded structures and thus are more likely to be accessible by the RISC/siRNA complex.

In addition to testing a variety of siRNAs targeting to different regions of the Bcl-2 and cyclin B1 genes, we have also tested our hypothesis by re-analysing the data of an independent study by Holen et al., who had examined the effects of a large number of siRNAs targeted to different regions of the hTF gene [10]. We found a very significant correlation between the H-b index and the siRNA efficacy (Fig. 3). These results strongly suggest that the H-b index can be a useful indicator for predicting the gene silencing efficiency of siRNA.

Other than the H-b index, we also found that formation of a hairpin structure within the siRNA target region can greatly reduce the efficiency of the siRNA. As shown in Table 3 and Fig. 4A & B, low silencing efficiency was detected for most siRNAs targeting to sites with a hairpin structure. We think this is because such siRNA may tend to form a hairpin structure by itself and thus cannot be fully open. As a result, the RISC/siRNA complex will be less effective in binding with the complementary mRNA.

In summary, based on results of this work, we proposed two guidelines for selecting siRNA target sites for effective RNA interference: (1) It is preferable to choose a target region that has a low H-b index (ideally less than 25). (2) One should avoid target regions where the mRNA can form a hairpin structure. Besides these, one may also consider to stay away from target sites at either the 5' or 3' ends of the mRNA. Since proteins involved in translational regulation or mRNA processing may bind to these terminal regions, they could interfere with the siRNA-mRNA interaction.

**ACKNOWLEDGEMENTS**
This work was supported by the Research Grants Council of Hong Kong (HKUST6109/01M, HKUST6104/02M, and AoE) and HIA project of HKUST.